\documentstyle[prb,aps,preprint,eqsecnum,tighten]{revtex}
\newcommand{\kr}{$K_{\rho}$}
\newcommand{\krm}{K_{\rho}}
\newcommand{\ks}{$K_{\sigma}$}

\newcommand{\tmx}{$(TMTSF)_2X$}

\begin{document}
\draft
\title{Bounded Luttinger liquids as a universality class of quantum critical
behavior}
\author{Johannes Voit}
\address{Fakult\"{a}t f\"{u}r Physik, Albert Ludwigs-Universit\"{a}t,
Hermann-Herder-Str. 3, D-79104 Freiburg (Germany) \\
and Theoretische Physik 1, Universit\"{a}t Bayreuth, 
D-95440 Bayreuth (Germany)\footnote{Long-term address} }
\author{Yupeng Wang}
\address{Institute of Physics and Center for Condensed Matter Physics, Chinese
	Academy of Sciences, Beijing 100080, P. R. China}
\author{Marco Grioni}
\address{Institut de Physique Appliqu\'{e}e, EPFL Lausanne, CH-1015 Lausanne,
Switzerland}

\date{\today}
\maketitle
\begin{abstract}
We show that 
one-dimensional quantum 
systems with gapless degrees of freedom and open boundary conditions form
a new universality class of quantum critical behavior, which we propose
to call ``bounded Luttinger liquids''. They share the following properties
with ordinary (periodic) Luttinger liquids: absence of fermionic 
quasi-particle excitations, charge-spin separation, anomalous power-law
correlations with exponents whose scaling relations are parametrized by
a single coupling constant per degree of freedom, $K_{\nu}$. The values
of $K_{\nu}$ are independent of boundary conditions, but the representation
of the critical exponents in terms of these $K_{\nu}$ depends on boundary
conditions. We illustrate these scaling relations by exploring general
rules for boundary critical exponents derived earlier using the Bethe 
Ansatz solution of the 1D Hubbard model together with boundary conformal
field theory, and the theory of Luttinger liquids in finite-size systems.
We apply this theory to the photoemission properties of the organic 
conductors $(TMTSF)_2X$, and discuss to what extent
the assumption of finite strands
with open boundaries at the sample surface can reconcile the experimental
results with independent information on the Luttinger liquid state in these
materials.
\end{abstract}

\narrowtext

\section{Motivation}
\label{moti}
Phase transitions take place in a different way on surfaces and in the bulk of
a sample \cite{11}. 
Order parameters, critical temperatures, critical exponents, and their 
scaling relations may be different, i.e. entirely new universality classes
may be realized. Moreover, one may observe new phenomena due to 
anisotropy and the breaking of translational invariance caused by the 
boundary, like oscillations 
in correlation functions which, in the bulk, are monotonous, coordinate 
dependences, and
in particular Friedel oscillations, in local quantities, etc. 

Here, we discuss such boundary critical phenomena in 
one-dimensional (1D) strongly correlated electron systems which possess,
in the bulk, a quantum critical point at zero temperature. 
Conformal invariance \cite{BPZ}, a 
consequence of the combined Lorentz and scale 
invariance at a critical point, allows an exact determination of 
all critical exponents  in 1+1D. The conformal field theory is parametrized 
by a unique constant -- the conformal anomaly or the central charge $c$
of the corresponding Virasoro algebra \cite{BPZ}. Strongly correlated 
electron systems of the kind we are interested in here, have $c=1$,
and their critical exponents continuously depend on the 
coupling of the fields. They can be calculated exactly from the
low-energy excitations of the underlying Hamiltonians 
\cite{ci,fk,myreview}.
The excitation spectra, in turn, can be obtained
by a variety of methods, notably Bethe Ansatz for integrable models,
and exact numerical diagonalization on small lattices quite generally.

Conformally invariant, 1D strongly correlated electron systems are 
metals, and are also 
described as Luttinger liquids \cite{myreview,Haldane}. In that perspective, 
the main focus is on their non-Fermi liquid properties which are embodied
in their critical exponents (anomalous dimensions) and different velocities
for excitations in different conformally invariant sectors (for two
sectors only, charge and spin, this leads to charge-spin separation). In
Luttinger liquid theory, the scaling relations between the different
critical exponents are parametrized by a
single renormalized coupling constant per degree of freedom 
$K_{\nu}$  (here $\nu = \rho$, charge,
and $\nu = \sigma$, spin), playing the role of the Landau parameters familiar
from Fermi liquid theory. The ``Luttinger liquid universality class'' then
corresponds to unique dependences of all critical exponents on the $K_{\nu}$.
The connection between both methods is well established in bulk systems
\cite{myreview,map,hjs}.

Recently, both approaches have been extended to boundary critical
phenomena in 1D fermion systems with open boundary conditions.
Boundary conformal field theory \cite{bcft} has been used to
derive the boundary critical exponents of integrable 1D electron systems
in terms of the dressed charge matrices of these models
\cite{wang,fuji}. (Following the early work of Gaudin \cite{gau}, Bethe Ansatz
solutions have been produced, and integrability proven, for certain
1D electron models with boundaries \cite{rba,boun}.)
Also Luttinger liquid theory has been formulated for 
systems with open boundaries \cite{fabgog,johan,matt}. 
While Wang \em et al.\rm \cite{wang} gave the rules for evaluating 
critical exponents, explicit expressions for the exponents of 
specific correlation functions, relevant values for specific models  
such as the 1D Hubbard model, and their relation to those derived
within Luttinger liquid theory, are still missing. An exception are Friedel
oscillations in the 1D Hubbard model which have been
studied by Bed\"{u}rftig \em et al., \rm \cite{bed}. Here, and for spinless
fermions \cite{schmitteck}, accurate DMRG calculations are in impressive
agreement with predictions from conformal field theory \cite{wang}. 
Related problems have also been studied in the context of quantum spin
chains \cite{affegg}.

Here, we provide explicit expressions for the critical exponents of
a variety of correlation functions,
and explicitly connect the predictions of conformal field theory in terms
of the dressed charge matrices of integrable models to standard Luttinger
liquid notation. We discuss the consequences of such mappings for nonintegrable
systems and for experiments. We discuss the consequences of broken 
translational invariance in open systems.
To be specific, we discuss the 1D Hubbard model throughout this paper,
except when stated otherwise. The results can be carried over to other
integrable models without difficulty. 

Quite generally, we are interested in how quantum critical
1D strongly correlated electron systems fit into the general framework
of surface critical phenomena, and of different universality classes
in the bulk and at surfaces. We find that the boundary critical exponents
of gapless (conformally invariant) 1D electron systems define a new
universality class which we propose to call \em bounded Luttinger liquids,
\rm and which affect almost all physical quantities.

We are motivated by several recent developments, both theoretical
and experimental. Recent theoretical work on boundary effects in
photoemission emphasizes the need for exact information on the spectral
function of 1D Hubbard models with open boundaries. Approximate results
would indicate an enhancement of spectral weight in a large energy
range around the chemical potential \cite{bcesch}. 
On the other hand, unexplained
photoemission experiments on the quasi-1D Bechgaard salts \cite{phot,zwick}
have been
tentatively associated with the possible influence of boundary effects
\cite{johan,grioni}. Finally, boundary critical exponents are relevant
in mesoscopic quantum wires and quantum Hall edge states 
and carbon nanotubes \cite{kafi,rother,bockrath}, 
and the Luttinger liquid language
has been preferred in all these articles.

As we shall see, integrable models can be used to illustrate 
relations between bulk and boundary critical exponents which are
valid beyond the realm of integrability, in non-integrable Hamiltonians
and even in experimental Luttinger liquids, provided they exist.
Remarkably, they allow to propose boundary 
effects as a possible resolution of the qualitative discrepancy of the
photoemission spectra of the 1D Bechgaard salts, and the information 
provided by almost all other experiments on these materials. 

The reminder of this paper is organized as follows. 
In Section \ref{lemodel} we discuss the determination of critical exponents
of 1D correlated electron systems with periodic and open boundary conditions.
We show which quantities determining the critical exponents
are independent of boundary conditions, and which
do depend on them. We do this for the 1D Hubbard model where we discuss
the results from the Bethe Ansatz solution and use boundary conformal field
theory, and for Luttinger liquids. For Luttinger liquids, we rewrite all
critical exponents in a form similar to conformal field theory, i.e. in
terms of the coupling constants $K_{\nu}$, and the quantum numbers which
an operator inserts into a system. The rules derived here are used in 
Section \ref{bll} to give explicit expressions for a variety of correlation
functions of bounded Luttinger liquids, and to illustrate the new scaling
relations found here. Section \ref{appl} discusses the application of 
boundary critical phenomena to problems posed by photoemission experiments
on a class on 1D organic conductors, and Section \ref{oq} summarizes the
open questions remaining. 

\section{Models}
\label{lemodel}
While our results are valid generally for interacting 1D electron systems,
we shall concentrate our discussion on the 1D Hubbard model as a prototypical
example of an integrable (or Bethe ansatz solvable, used synonymously) 
system, and on the 1D Luttinger model. Here, we briefly describe how their
critical properties are calculated, and how the solutions with periodic and
open boundary conditions are related. 

\subsection{The 1D Hubbard model}
\label{modhub}
The 1D Hubbard model is defined by the Hamiltonian
\begin{eqnarray}
\label{hubham}
H=-\sum_{i=1}^{N-1}\sum_{\sigma=\pm}\left(C_{i\sigma}^\dagger C_{i+1\sigma}
+ {\rm H. c.} \right) +
U\sum_{i=1}^{N}n_{i\uparrow}n_{i\downarrow}
-\mu\sum_{i=1}^{N}\sum_{\sigma=\pm}n_{i\sigma}-
\frac h2\sum_{i=1}^{N}(n_{i\uparrow}-n_{i\downarrow}),
\end{eqnarray}
where $C_{i\sigma}$ ($C_{i\sigma}^\dagger$) is the electron annihilation 
(creation) operator;
$\mu$ denotes the chemical potential and $h$ is the external magnetic 
field. We shall consider $h=0$ in much of, and non-half-filled bands
throughout this paper. Eq.\ (\ref{hubham}) represents the Hamiltonian on
$N$ sites with open boundary conditions (OBC). For periodic boundary
conditions (PBC), extend the first sum to $N$ and identify $C_{N+1} = C_1$.

This model is solved by Bethe Ansatz \cite{ba} for periodic and by
reflection Bethe Ansatz for open boundary conditions \cite{rba}, 
and the energies of the
ground and excited states are given as the solution of certain integral
equations. Here we do not repeat these equations from the
literature but emphasize the main physical results on passing from periodic
to open boundary conditions. The basic idea of the reflection Bethe Ansatz
is to superpose, in order to obtain the solution for open boundaries, 
the Bethe solution of the periodic system with its reflection at the origin
\cite{gau}. In a similar way, one can pass from a particle on a ring
to a particle in a box in elementary quantum mechanics, from a 
Luttinger liquid with periodic boundary conditions to one with open
boundaries \cite{fabgog}, and from  conformal to boundary conformal field
theory \cite{bcft}. A many-particle eigenstate with $N_c = 
N_{\uparrow} + N_{\downarrow}$ electrons out of which $N_s = N_{\downarrow}$
have spin projection -1/2, is parametrized by the
set of rapidities ($k$, $\Lambda$ for charge and down-spin, basically a
generalization of the wavenumbers of free electron states to include the
correct scattering phase shifts) of the occupied 
states \cite{notation}. While
for PBC, rapidites $+k$ and $-k$ give linearly independent solutions, for
open boundaries only one sign (say $+$) for these rapidites is allowed
(as in elementary quantum mechanics), but the spacing of solutions is half
as big and their density twice as big as for PBC. 
The equations determining the energies are formally different for 
open and periodic boundary conditions. However, they can be transformed
into each other (cf. below for the entries of the dressed charge matrix),
and consequently, their solutions are equal. The energies
thus are independent of the boundary conditions, as are
the velocities of the low-lying collective charge and spin modes
(holons and spinons). 

The critical
properties are determined by the low-energy excitations, more specifically
by the quantum numbers associated with the operators whose correlations at
criticality we wish to determine, and a ``dressed charge matrix'' $Z$
\cite{fk,wang}. 
This $2 \times 2$-matrix
contains the effective renormalized coupling constants within and between
the low-energy (charge and spin) sectors of the Hilbert space of the 
Hubbard model, and therefore directly 
determines the non-universal critical exponents. It is defined as
\begin{equation}
\label{dcm}
Z^{(p,b)} \equiv \left( 
\begin{array}{cc}
Z_{cc} & Z_{cs} \\
Z_{sc} & Z_{ss} 
\end{array}
\right) 
\; .
\end{equation}
The superscripts $p,b$ label periodic and bounded
systems. In the absence of a magnetic field, $Z^{(p,b)}$ is completely
determined by its first element $Z_{cc}^{(p,b)}$ as
\begin{equation}
\label{dcmh0}
Z^{(p,b)} = \left( 
\begin{array}{cc}
\xi^{(p,b)}(k_0) & 0 \\
\frac{\xi^{(p,b)}(k_0)}{2} & \frac{1}{\sqrt{2}}
\end{array}
\right) 
\end{equation}
where $k_0$ is a cutoff determined by the particle 
density, again identical for periodic and open
boundary conditions. 

$\xi^{(p,b)}(k_0)$, for a periodic system, obeys the integral equation
\cite{fk}
\begin{equation}
\label{ixip}
\xi^{(p)}(k) = 1 + \frac{1}{2 \pi} \int_{-k_0}^{k_0} \! dk' \cos(k') 
K^{(p)} (\sin k - \sin k') \xi^{(p)}(k') 
\end{equation}
with the kernel
\begin{equation}
\label{kp}
K^{(p)}(z) = \int_0^{\infty} \! d \omega \frac{\exp(- \omega U / 4)}{
\cosh(\omega U /4)} \cos (\omega z) \;.
\end{equation}
For open boundary conditions, on the other hand, the integral equation
for $\xi^{(b)}(k_0)$ reads 
\begin{equation}
\label{ixio}
\xi^{(b)}(k) = 1 + \frac{1}{2 \pi} \int_{0}^{k_0} \! dk' \cos(k') 
K^{(b)} (\sin k , \sin k') \xi^{(b)}(k') 
\end{equation}
with the kernel 
\begin{equation}
\label{ko}
K^{(b)}(z,z') = K^{(p)}(z-z') + K^{(p)}(z+z') \; .
\end{equation}

Using $ K^{(b)}(-z,z') = K^{(b)}(z,z')$, a consequence of $ K^{(p)}(-z) 
= K^{(p)}(z)$, Eq.\ (\ref{kp}), we find that $\xi^{(b)}(k) $
can be continued to negative $k$ with $\xi^{(b)}(-k)  = \xi^{(b)}(k)$.
We can then change variables $k \rightarrow -k$ in the contribution of
the second term of $K^{(b)}$ to the integral in (\ref{ixio}), and find
\begin{equation}
\xi^{(b)}(k) = \xi^{(p)}(k) \; {\rm and} \;
{\rm thus} \; Z^{(b)} = Z^{(p)} \;.
\end{equation}
Despite the differences between Eqs. (\ref{ixip}), (\ref{kp}), and
(\ref{ixio}), (\ref{ko}), 
the dressed charge matrices are identical for open and periodic boundary
conditions. This translates the fact that they are a property of the 
Hamiltonian only and independent of boundary conditions. 
The boundary effects are contained completely in the representation
of the boundary critical exponents in terms of the 
entries of the dressed charge matrices. This representation
depends on the boundary conditions. 

With a magnetic field, one has a set of four coupled equations for the
four entries of $Z^{(p,b)}$, containing different kernels. However,
all symmetries used, and all transformations carried out above continue 
to be applicable. Consequently, the dressed charge matrices 
of open and periodic systems are identical also in a finite magnetic 
field. 

The rapidities $k, \Lambda$, describing a general quantum state of the 
1D Hubbard model depend on two 
sets of quantum numbers $I_{c,j}$ and $I_{s,j}$ of integers or half-odd 
integers, parametrizing the solutions (the ``occupied states'').
There are certain parity rules for the $I_{c,j}$, $I_{s,j}$,
depending on $N_{c}$, $N_s$ being even or odd. In the ground state, their
distributions are filled up to a cutoff, a pseudo-Fermi number.
Acting with an operator $O$ on a state of the system will change the
distribution of rapidities. Both for the charge and spin channels,
in the periodic systems, there are three types of excitations:
particle addition (often also termed ``charge" excitations \cite{charge}), 
current
excitations, and particle-hole excitations. The single-particle operator
$C_{n,s}^{\dagger}$, e.g., adds a particle, i.e. $\Delta N_c = 1$,
and, depending on spin, $\Delta N_s = 0$ for $s= \uparrow$ or 
$\Delta N_s = 1$ for $s= \downarrow$. Changing  $N_{c,s}$ by unity 
changes the 
$I_{c,j}$ and/or $I_{s,j}$ between integers and half-odd integers. This
backflow is accounted for by current quantum numbers $D_{c}
= (\Delta N_{c}+\Delta N_s)/2 $ mod 1, and $D_{s}
= \Delta N_s/2 $ mod 1. The role of these excitations, and of the
particle-hole excitations, can best be seen by considering, e.g. the
density operator $\sum_s c^{\dagger}_{n,s} c_{n,s} = N^{-1} \sum_{k,q,s}
c^{\dagger}_{k+q,s} c_{k,s}$. This operator does not change the particle
number, i.e. $\Delta N_c =  \Delta N_s = 0$. For small $q$, this operator
creates particle-hole excitations, i.e. changes the distribution of the 
positive (or negative) $I_{c,j}$ or $I_{s,j}$ by some $\Delta I_c$ or
$\Delta I_s$. If $q$ is not small and rather a multiple of $2k_F$, 
particles are transferred across the Fermi surface which will generate
a persistent current in the system. These large-$q$ excitations therefore
change the symmetry of the distribution of the $I_{c,j}$ and $I_{s,j}$,
and therefore have a finite $D_{c,s}$. Permissible values of the $D_{c,s}$
follow from the expressions above. The finite $D_{c,s}$ values then
generate the $2k_F$, $4k_F$, etc. excitations. 

This set of quantum numbers is characteristic for an operator $O(x)$.
Conformal field theory then determines its correlation functions
as \cite{notation}
\begin{equation}
\label{cfop}
\langle O(xt) O^{\dagger} (00) \rangle \sim  
\frac{ a(D_c, D_s) \exp(- 2i D_c k_{F \uparrow} x) \exp(- 2i [D_c+D_s]
k_{F \downarrow} x)}{(x - v_{\rho} t)^{2 \Delta_c^+} 
(x + v_{\rho} t)^{2 \Delta_c^-} 
(x - v_{\sigma} t)^{2 \Delta_s^+} 
(x + v_{\sigma} t)^{2 \Delta_s^-} } \;.
\end{equation}
The anomalous dimensions of the operator 
\begin{eqnarray}
\label{bulkconch}
\Delta_c^{\pm} & = & \frac{1}{2} \left( \pm \frac{Z_{ss} \Delta N_c -
Z_{cc} \Delta N_s}{ 2 \det Z} + Z_{cc} D_c + Z_{sc} D_s
\right)^2 + \Delta I_c^{\pm}\\
\label{bulkcons}
\Delta_s^{\pm} & = & \frac{1}{2} \left( \mp \frac{Z_{sc} \Delta N_c -
Z_{cc} \Delta N_s}{ 2 \det Z} + Z_{cs} D_c + Z_{ss} D_s
\right)^2 + \Delta I_s^{\pm}
\end{eqnarray}
are determined by the entries of the dressed charge matrix and 
the changes in the quantum numbers $\Delta N_{c,s}$ and $D_{c,s}$ 
which the action of the operator generates.

For bounded systems, only positive rapidities are relevant, 
cf. Eq.\ (\ref{ixio}), there is only one pseudo-Fermi point for each channel,
and there are no current excitations. 
Consequently, there are only charge excitations $\Delta N_c$, 
$\Delta N_s$, and particle-hole excitations  $\Delta I_c$, 
$\Delta I_s$.
The backflow terms, resp. asymmetries of the rapidity
distributions, $D_{c}$, $D_s$ are absent. 

The complete, explicit behavior of correlation functions is rather complicated
close to an open boundary, cf. Eq. (8) of Wang \em et al. \rm \cite{wang},
for an example for a single channel (charge or spin). Asymptotically, 
however, the correlation functions $G_{O}$ of an operator
$O$ simplify considerably and behave, close to an open boundary at $x=0$,  
\begin{enumerate}
\item in time (for $x_1, x_2 \approx 0$, and $v t \gg x_1, x_2$) as
\begin{equation}
\label{gtime}
G_{O}(x_1, x_2, t) \sim t^{-2 x_{O,b}}
\end{equation}
\item in space (specifically: $x_1, t \approx 0, x_2 \gg
x_1, v t$) as
\begin{equation}
\label{gspace}
G_{O}(x_1, x_2, t) \sim x_2^{-(x_{O,b} + d_{O})} \;.
\end{equation}
(Only the conformally invariant, i.e. slowly varying part of the correlation
function is given. This must be multiplied by a factor oscillating
with an appropriate multiple of $k_F x_2$.) 
\item in temperature ($x_1, x_2 \approx 0, \omega \ll T$)
\begin{equation}
\label{gtemp}
G_{O}(x_1, x_2, \omega, T) \sim T^{2 x_{O,b} -1} \;,
\end{equation}
where the Fourier-transformed (in time) correlation function is used.
That the boundary critical exponent $x_{O,b}$, describing temporal 
correlations, comes in for temperature is most easily rationalized 
from the perspective of the Matsubara formalism of imaginary times
(frequencies) in many-body physics.
\end{enumerate}
The exponents $x_{O,b}$ and $d_{O}$ 
are the boundary critical
exponent and bulk scaling dimension, 
respectively. The different exponents for spatial and time decays,
and the appearence of a new boundary critical exponent
describing the time correlations, translate the 
combined influences of the interactions and of 
broken translational invariance, on the low-energy properties of interacting
1D electron systems close to boundaries.
In terms of the dressed charge matrix $Z$, and
the quantum numbers $\Delta N_c$, $\Delta N_s$, $\Delta I_c$, and
$\Delta I_s$ associated with the operator $O$, $x_{O,b}$
is given by \cite{wang}
\begin{equation}
\label{xbwang}
x_{O,b} = \frac{1}{2}
(Z^{-1} {\bf \Delta N})^T \cdot (Z^{-1} {\bf \Delta N}) 
+ \Delta I_c + \Delta I_s 
\end{equation}
with $({\bf \Delta N})^T = (\Delta N_c, \Delta N_s)$, and the superscript $T$
denotes the transpose. The bulk critical exponent is 
\begin{equation}
d_O = \Delta_c^+ + \Delta_c^- + \Delta_s^+ + \Delta_s^- \;.
\end{equation}
Finally notice that when the system has boundaries on both sides, i.e.
is of finite length $L$, the power laws of space and time are changed
to power laws of $\sin(\pi x / 2L)$, resp. $\sin(\pi vt / 2L)$ with the
same exponents.

\subsection{Luttinger liquid theory}
At low energies, models of correlated 1D electrons such as the 
1D Hubbard model, reduce to Luttinger liquids \cite{myreview,Haldane}. 
Haldane's Luttinger liquid conjecture \cite{Haldane} postulates that,
in a low-energy subspace, gapless one-dimensional quantum systems 
are described by an effective Luttinger Hamiltonian \cite{notation}
\begin{equation}
H  =  \frac{1}{2 \pi} \sum_{\nu= \rho,\sigma} 
\int \! dx \left\{ v_{\nu} K_{\nu} \: \pi^2
\Pi_{\nu}^2(x) + \frac{v_{\nu}}{K_{\nu}} \left( \frac{\partial 
\Phi_{\nu}(x)}{\partial x} \right)^2 \right\}  \;\;.
\end{equation}
$\Phi_{\nu}$ and $\Pi_{\nu}$ are phase fields and their conjugate momenta
for charge ($\nu  = \rho$) and spin ($\nu = \sigma$) fluctuations
\begin{equation}
\label{phi}
\Phi_{\nu}(x) = - \frac{i\pi}{L} \sum_{p \neq 0 } 
\frac{e^{-\alpha \mid p \mid / 2 -ipx}}{p} \left[\nu_+(p)+\nu_-(p) \right]
\;\;\;,
\end{equation}
and
\begin{equation}
\label{theta}
\Pi_{\nu}(x) = \frac{1}{\pi} \frac{\partial \Theta_{\nu}(x)}{\partial x}
\;\; {\rm with} \;\;
\Theta_{\nu}(x) = \frac{i\pi}{L} \sum_{p \neq 0 } \frac{e^{-\alpha
\mid p \mid / 2 -ipx}}{p} \left[\nu_+(p) - \nu_-(p) \right] 
\;\;\;,
\end{equation}
which obey bosonic commutator relations,  as do the long-wavelength
fluctuation operators $\nu_r(p)$. The $K_{\nu}$ are the renormalized
coupling constants in the charge and spin sectors. Zero-modes ($p=0$)
do not influence the dynamics, and have been neglected. 

Correlation functions and their critical exponents can then be
calculated by bosonization \cite{Haldane,myreview}
\begin{equation}
\label{bos}
\Psi_{rs}(x) \sim \lim_{\alpha \rightarrow 0}
\frac{ e^{irk_Fx} }{ \sqrt{2 \pi \alpha} } \exp \left( \frac{-i}{\sqrt{2}} 
\left[ r \Phi_{\rho}(x) - \Theta_{\rho}(x)
+ s \left\{ r \Phi_{\sigma}(x) - \Theta_{\sigma}(x) \right\} 
\right] \right) \;\;\;.
\end{equation}
The chirality label $r=\pm$ here and in Eqs.\ (\ref{phi}) and (\ref{theta}),
refers to right- and left-moving fermions, with $k \approx r k_F$. 
Consider a general local operator
\begin{equation}
\label{llop}
O_{ \{m\}}(x) \equiv O_{m_{+ \uparrow}, m_{+ \downarrow}, m_{- \uparrow}
m_{- \downarrow}}(x) = \Psi_{+ \uparrow}^{m_{+ \uparrow}}(x)
\Psi_{+ \downarrow}^{m_{+ \downarrow}}(x) 
\Psi_{- \uparrow}^{m_{- \uparrow}}(x) 
\Psi_{- \downarrow}^{m_{- \downarrow}}(x) \;\;,
\end{equation}
where positive (negative) exponents label powers of creation 
(annihilation) operators, and higher powers of the operators are understood
to be point-split, e.g., ($m_{+ \uparrow} > 1$)
\begin{equation}
\Psi_{+ \uparrow}^{m_{+ \uparrow}}(x) = \prod_{i=1}^{m_{+ \uparrow}}
\Psi_{+ \uparrow}^{\dagger} (x+[i-1]a) \;\;.
\end{equation}
The two-point correlation function of such an operator decays as
\begin{equation}
\label{cfo}
\langle O_{ \{m\}}(xt) O^{\dag}_{ \{m\}}(00) \rangle \sim 
\frac{\exp ( - i k_F x J_{\rho})}{
(x - v_{\rho}t)^{2 d_{\rho}^+} (x + v_{\rho}t)^{2 d_{\rho}^-}
(x - v_{\sigma}t)^{2 d_{\sigma}^+} (x + v_{\sigma}t)^{2 d_{\sigma}^-}}
\end{equation}
where the scaling dimensions 
\begin{eqnarray}
2 d_{\rho}^+ & = & \frac{1}{8} \left[ (\Delta N_{\rho} + J_{\rho} )^2
+ (K_{\rho} - 1) J_{\rho} + (K_{\rho}^{-1} - 1) \Delta N_{\rho} \right] 
\nonumber \\
2 d_{\rho}^- & = & \frac{1}{8} \left[ (\Delta N_{\rho} - J_{\rho} )^2
+ (K_{\rho} - 1) J_{\rho} + (K_{\rho}^{-1} - 1) \Delta N_{\rho} \right] 
\nonumber \\
\label{scadim}
~ & ~ & ~ \\
2 d_{\sigma}^+ & = & \frac{1}{8} \left[ (\Delta N_{\sigma} + J_{\sigma} )^2
+ (K_{\sigma} - 1) J_{\sigma} + (K_{\sigma}^{-1} - 1) \Delta N_{\sigma} 
\right]  \nonumber\\
2 d_{\sigma}^- & = & \frac{1}{8} \left[ (\Delta N_{\sigma} - J_{\sigma} )^2
+ (K_{\sigma} - 1) J_{\sigma} + (K_{\sigma}^{-1} - 1) 
\Delta N_{\sigma} \right] \nonumber
\end{eqnarray}
are determined by the two Luttinger liquid parameters $K_{\nu}$ and the
number of charge and current excitations in the charge and spin sectors,
created by the operator $O_{ \{ m \} }(xt)$
\begin{equation}
\label{quanu}
\Delta N_{\rho} = \sum_{r,s} m_{r,s} \;, \quad
J_{\rho} = \sum_{r,s} r m_{r,s} \;, \quad
\Delta N_{\sigma} = \sum_{r,s} s m_{r,s} \;, \quad
J_{\sigma} = \sum_{r,s} rs m_{r,s} \;.
\end{equation}
In addition, there are particle-hole excitations 
with chirality $r$ in each sector, given
by 
\begin{equation}
\label{phexx}
\rho_r(x) = \sum_s \Psi_{rs}^{\dag}(x) \Psi_{rs}(x) \;, \quad
\sigma_r(x) = \sum_{s} s \Psi_{rs}^{\dag}(x) \Psi_{rs}(x) \;.
\end{equation}
If an operator $\tilde{O}$ differs from $O$ above by $\Delta I_{r,\nu}$
powers of $\nu_r(x)$-type particle-hole excitations, the exponent of 
its correlation function will be increased as
\begin{equation}
\label{phexdim}
d_{\nu}^r \rightarrow d_{\nu}^r + \Delta I_{r, \nu} \;.
\end{equation}
Up to now, these critical exponents have been derived case by case, by
an explicit bosonization calculation. Our Eqs.\ (\ref{cfo}) and (\ref{scadim})
show that, in complete analogy with conformal field theory, 
it is possible to give a general construction rule for the correlation
functions, only based on the knowledge of the $K_{\nu}$ and the 
quantum numbers associated with the operators.

This picture is well-established for periodic systems \cite{myreview}.
For open systems, the same basic ideas as outlined above, for integrable
models and boundary conformal field theory, continue to hold:
as a consequence of the boundary conditons, right- and left-moving fermion
fields are no longer independent, 
\begin{equation}
\label{psib}
\Psi_{+,s}(x) = - \Psi_{-,s}(-x) \;,
\end{equation}
and the physical
fermions are described as superpositions of a chiral fermion with
its reflection at the origin \cite{fabgog,johan,matt}. 
The basic bosonization
formula (\ref{bos}) is then replaced by
\begin{equation}
\label{bos:obc}
\Psi_{+,s}(x) \sim \lim_{\alpha \rightarrow 0}
\frac{ e^{ik_Fx} }{ \sqrt{2 \pi \alpha} } \exp \left( \frac{-i}{\sqrt{2}} 
\left[ \Phi_{+\rho}(x) + s \Phi_{+\sigma}(x)  
\right] \right) \;\;\;,
\end{equation}
where the fields $\Phi_{+,\nu}(x)$ are obtained from Eqs.\ (\ref{phi})
or (\ref{theta}) by dropping the $\nu_{-}(p)$-operators. Due to the
open boundary conditions, the full correlation functions of these operators
are quite involved. While the single-particle Green's function has 
been calculated by others \cite{fabgog,johan,matt}, 
general correlation functions
have not been published to date. Here, we give simple rules how the 
exponents describing the 
asymptotic decay of the correlation functions of more complicated operators,
close to an open boundary, can be constructed in complete analogy to 
the recipes of boundary conformal field theory. 

In the presence of open boundaries, the general operator
$O_{\{m\}}(x)$ is represented in terms of only one type of chiral fermion,
say $r=+$. There is thus only a single Fermi point, and no current excitations
can be defined. There are only charge and particle-hole excitations.
We therefore have $O_{\{m\}}^{(p)} \rightarrow O_{\{m\}}^{(b)}$ where the
superscripts refer to boundary conditions 
\begin{equation}
\label{llop:obc}
O_{\{m\}}^{(b)}(x) = \Psi_{+\uparrow}^{m_{+\uparrow}+m_{-\uparrow}}(x)
\Psi_{+\downarrow}^{m_{+\downarrow}+m_{-\downarrow}}(x) =
\Psi_{+\uparrow}^{\Delta N_{\uparrow}}(x)
\Psi_{+\downarrow}^{\Delta N_{\downarrow}}(x)
\end{equation}
with the same conventions as for $O_{\{m\}}^{(p)}$, Eq.\ (\ref{llop}).
The first equality directly translates the bulk operator, in terms of its
bulk quantum numbers, into a boundary operator, while the second equality
only uses the quantum numbers defined for the boundary operator:
$\Delta N_s$ is the number of spin-s particles the operator $O$ adds to
the system. One now can calculate the correlations functions of 
$O_{\{m\}}^{(b)}(x)$ by the methods developed by others
\cite{fabgog,johan,matt}. 

Close to an open boundary in a semi-infinite system, the complicated 
expressions simplify considerably in the limits $t \gg x$ and $x \gg t$
where $x$ is close to the boundary. From these limits, a boundary
critical exponent $x_{O,b}$ can be defined for each operator $O_{\{m\}}(xt)$,
and the behavior of the correlation functions as functions of the variables
$x$, $t$, and $T$ is formally identical to 
Eqs.\ (\ref{gtime})--(\ref{gtemp}). 
For a Luttinger liquid, the boundary critical exponent 
$x_{O,b}$ then is given by
\begin{equation}
\label{bcelutt}
2 x_{O,b}  =  x_{O,b}^{(\rho)} +  x_{O,b}^{(\rho)} \;, \quad
x_{O,b}^{(\nu)}  =  \frac{(\Delta N_{\nu})^2}{4  K_{\nu}} \;.
\end{equation}
Formally, $\Delta N_{\nu}$ is calculated in the same way as above,
Eqs. (\ref{quanu}), when the operator is represented in terms of the 
bulk chiral fermions, resp. with the correct number of $m_{+,s}$ and
$m_{-,s} \equiv 0$ when it is set up directly as a boundary operator. 
Physically, an operator which generated current excitations in a periodic
system, will now generate particle-hole excitations. 
The spatial decay of the correlations of $O(x)$ close to a boundary
involves the bulk scaling dimension
$d_O = \sum_{r,\nu} d_{\nu}^r$ 
of the operator $O(x)$ as in (\ref{gspace}), and the $d_{\nu}^r$
are taken from Eqs.\ (\ref{scadim}). The temperature variation again
is determined by the boundary critical exponent $x_{O,b}$ as in (\ref{gtemp}).
Bosonization of a Luttinger liquid with open boundaries therefore precisely
reproduces the structure of correlation functions, and the boundary critical
exponents that boundary conformal field theory extracts for integrable 
systems. 

Notice further that for Luttinger liquids with open boundaries, the
value of $K_{\rho}$ again comes out identical to that for periodic systems,
translating the fact that it is a property of the Hamiltonian, resp.
thermodynamics of the system, and as such independent of boundary conditions.
Besides determining the boundary critical exponents of 1D fermions with
open boundaries, our work therefore provides rules, in terms of \kr ,
how to connect bulk and boundary critical exponents of different correlation
functions. Below, we shall discuss
an application of this procedure. 

Both methods, boundary conformal field theory and the Luttinger liquid,
must lead to identical critical exponents. The relation between the 
entries of the dressed charge matrix, and the Luttinger liquid $K_{\rho}$
in the absence of magnetic fields (then
$K_{\sigma} \equiv 1$ for spin-rotation invariance) is
\begin{equation}
\label{ident}
\xi^2(k_0) = 2 K_{\rho} \;,
\end{equation} 
both for bounded and for periodic systems. 

In a finite magnetic field, the situation is far more complicated. 
In the 1D Hubbard model, there are (at least) three  important effects
of a finite magnetic field $h$ \cite{fkmag}:
\begin{enumerate} 
\item \em All \rm entries of the dressed charge matrix acquire $h$-dependent
corrections. This implies that 
the Luttinger coupling constant $K_{\rho}$, to the extent that it still
makes sense, would depend on $h$ and therefore that charge-spin separation
is violated, even in a low-energy subspace. Explicit expressions have been 
given by Frahm and Korepin
\cite{fkmag}. Specifically, $Z_{cs} \neq 0$ now. 
\item The correction to the matrix element $Z_{ss}$ is particular in that
it is logarithmic in $h$
\begin{equation}
\label{zssmag}
Z_{ss} = \frac{1}{\sqrt{2}} \left( 1 + \frac{1}{4 \ln (h_c / h)} \right) \;,
\end{equation}
while all other corrections have weaker $h$-dependences.
$h_c$ is the critical field for a completely spin-polarized state. 
For 
the 1D Hubbard model, it has been calculated by Frahm and Korepin \cite{fkmag}
and varies from a finite constant at $U=0$ to a $1/U$ behavior as $U
\rightarrow \infty$. For non-interacting 1D electrons, one has $h_c =
\varepsilon(2k_F)$ where $k_F$ is the Fermi wavevector for $h=0$ and $
\varepsilon(k)$ is the single-particle dispersion taken from the bottom of 
the band, $\varepsilon(0) = 0$. Thus $h_c = 2t[1-\cos(2k_Fa)]$ for 
1D tight-binding electrons. 
\item There are different Fermi wave vectors for up- and down-spin electrons, 
$k_{F \uparrow} \neq k_{F \downarrow}$. 
\end{enumerate}
Quite generally, the mapping to a Luttinger liquid is problematic
and cannot be carried out any more. If one
is interested in boundary critical properties of 1D integrable systems, 
the conformal field theory approach (\ref{gtime})--(\ref{xbwang}) must
be preferred. 

In the (apparently realistic) limit $h \ll h_c$, the situation simplifies,
however, and a mapping to a Luttinger liquid can still be done \em
to logarithmic accuracy. \rm To this accuracy, the only effect of the 
magnetic field is the correction to $Z_{ss}$, Eq.\ (\ref{zssmag}). This 
translates in a $h$-dependent deviation from unity, of the Luttinger
coupling constant 
\begin{equation}
\label{ks}
K_{\sigma} = \left( 1 + \frac{1}{4 \ln (h_c/h)} \right)^2 
\approx 1 + \frac{1}{2 \ln (h_c/h)} \;.
\end{equation}
We obtained this equation  by evaluating the finite-size corrections
to the ground state energy \cite{fk},
\begin{equation}
E({\bf \Delta N}, {\bf D}) - E_0 = \frac{2 \pi}{N}
\left[ v_{\rho} (\Delta_c^+ + \Delta_c^-) + v_{\sigma} (\Delta_s^+ +
\Delta_s^-) \right]
\end{equation}
from states with finite $\Delta N_s$ but vanishing $\Delta N_c = D_c = D_s
= 0$, and using (\ref{bulkconch}) and (\ref{bulkcons})
\begin{equation}
E(\Delta N_s, 0,0,0) - E_0 = \frac{2 \pi}{N} v_{\sigma} 
\left( \frac{\Delta N_s}{2 Z_{ss}} \right)^2 \;.
\end{equation}
Its second derivative with respect to $\Delta N_s$ can then be identified
to the inverse susceptibility of a Luttinger liquid and gives (\ref{ks}).
To logarithmic accuracy in $h$, $K_{\rho}$ is unchanged, and one can
use also the expressions for the (boundary) critical exponents of Luttinger
liquids. A relation similar to (\ref{ks}) is also known for the 1D Heisenberg
model \cite{ksh}, and a correction to the magnetic susceptibility of the 1D
Hubbard model, logarithmic in $h$ and consistent with (\ref{ks}), has been
derived by Kawano and Takahashi \cite{kt}.

\section{A new universality class: bounded Luttinger liquids}
\label{bll}
Universality classes of critical behavior are usually defined in terms
of the scaling relations between the critical exponents. From the 
general expressions (\ref{gtime})--(\ref{gtemp}) above, it is clear
that these scaling relations on the boundary of a Luttinger
liquid are different from those 
in the bulk. Luttinger liquids with open boundaries therefore
realize a new universality class of (boundary) critical behavior which
we propose to call \em bounded Luttinger liquids. \rm More specifically,
following Haldane's statement for the periodic systems \cite{Haldane}, 
we conjecture
that \em one-dimensional quantum systems with gapless degrees of freedom
and open boundaries form bounded Luttinger liquids. \rm They comprise 
interacting electron systems, of which we discuss an example here, but also
1D bosons or spin chains \cite{wang}. 
Bounded Luttinger liquids are boundary conformal 
field theories with central charge $c=1$ \cite{wang}. Their nonuniversal
exponents
depend on one open coupling constant per degree of freedom $K_{\nu}$
which is the same as in the bulk. 

Here, we evaluate explicitly a variety of important 
boundary correlation functions. This will display
the new set of scaling relations characterizing the bounded Luttinger
liquids, and thus back our claim of a new universality class. 

The single-electron Green's function (spin-$\uparrow$) is
\begin{equation}
G_{\Psi}(x_1,x_2,t)= \langle \Psi(x_1)_{\uparrow}(t)
\Psi(x_2)_{\uparrow}^\dagger(0)
\rangle 
\sim \left\{  \begin{array}{l}
t^{-2 x_{\Psi,b}} \\
\sin(k_F x) x^{-x_{\Psi,b} - d_{\Psi}}
\end{array}\right.
= \left\{  \begin{array}{l}
t^{-1 -  \alpha_{\Psi,b}^t} \\
\sin(k_F x) x^{-1-\alpha_{\Psi,b}^x}
\end{array} \right.
\end{equation}
where the last equality transforms our notation to the familiar Luttinger
liquid notation with Green's function exponents $\alpha$. It
involves the following excitations (both in terms of the Bethe ansatz for
the Hubbard model, and in terms of a Luttinger liquid)
\begin{eqnarray}
\Delta N_c & = & 1,{~~~}\Delta N_s=0, {~~~}\Delta I_c=\Delta I_s=0 \\
\Delta N_{\rho} & = & 1, {~~~}\Delta N_{\sigma} = 1, {~~~}\Delta 
I_{\nu} =0
\end{eqnarray}
In the absence of a magnetic field ($Z_{cs} = 0$), 
its boundary critical exponent is
\begin{equation}
2 x_{\Psi,b} = \frac{Z_{ss}^2 + Z_{sc}^2}{( \det Z)^2 }
\end{equation}
and 
\begin{equation}
x_{\Psi,b} + d_{\Psi} 
= \frac{1}{4}\left( 3 \frac{Z_{ss}^2 + Z_{sc}^2}{( \det Z)^2 }
+ Z_{cc}^2 + Z_{sc}^2 + Z_{ss}^2 - 2 Z_{cc}Z_{sc} \right) \;\;.
\end{equation}
Using (\ref{ident}), the Luttinger liquid notation obtains:
\begin{equation}
\label{altll}
\alpha_{\Psi,b}^t = \frac{1}{2 K_{\rho}} - \frac{1}{2} 
\end{equation}
and 
\begin{equation}
\alpha_{\Psi,b}^x 
= \frac{1}{8} \left( \frac{3}{K_{\rho}} + K_{\rho} - 4 \right) \;\;.
\end{equation}
The same expressions are obtained, of course, when using directly Eq.\
(\ref{bcelutt}).
Photoemission experiments measure the Fourier transform of $G_{\Psi}(t)$, 
the local spectral function 
\begin{equation}
\label{lspec}
\rho(x,\omega) = -\frac{1}{\pi} {\rm Im} G(x,\omega + \mu) 
\sim | \omega |^{\alpha_{\Psi,b}^t}\;\;,
\end{equation}
where $\mu$ is the chemical potential.
For repulsive interactions $K_{\rho} < 1$, we have
\begin{equation}
\label{ineqexp}
\alpha_{\Psi,b}^t > \alpha_{\Psi,p} = \frac{1}{4}
\left( \krm + \frac{1}{\krm } - 2 \right)
\end{equation}
which suggests that the pseudogap in the local density of states is
\em deepened \rm by the presence of boundaries. 
We will use these expressions in Section \ref{appl}
below to discuss photoemission experiments
in organic conductors. 

Due to its relevance to photoemission experiments, the conformal field
theory predicition (\ref{lspec}) has been checked for the 1D Hubbard
model.
No exact calculation of the entire spectral
function is available. Conformal field theory therefore provides the
only exact prediction of properties of the spectral function, though only
of its exponents. The question therefore is (i) to what extent other
methods produce exponents consistent with conformal field theory, and
(ii) if the deepening of the pseudogap in the local density of states
is found, too. These issues were addressed recently, 
using perturbative Hartree 
and numerical DMRG calculations\cite{bcesch}. A suprising
result of the Hartree calculations was an enhancement of the density of
states, instead of a suppression, close to the chemical potential
and close to the open boundary. DMRG
was not able to calculate the frequency dependence of $\rho(x, \omega)$.
For the zero-frequency matrix elements, it asymptotically verified, however,
a decay with system size as $L^{-\alpha_{\Psi,b}^t}$, as expected from
conformal field theory. Moreover, at finite $L$, the matrix element
approached the predicted power-law from below. Both facts suggest that
the depth of the pseudogap is increased by the presence of open boundary
conditions. However, that work also suggests a word of caution: for small
$U$, the asymptotic power-laws were approached only on very large lattices
($L > 100$ sites) \cite{bcesch}
implying very small energy scales for (\ref{lspec}) to be observed. 
How these scales depend on the interaction strength or range, and if
the pseudogap is over- or underestimated at smaller $L$ (the DMRG data
seem to suggest the second possibility \cite{bcesch}), deserves further
study. It could also be interesting to perform an exact calculation
of the entire spectral function at $U=\infty$ as has been done successfully
for periodic systems \cite{hallb}.

The boundary critical exponent of the Green's function also determines
the temperature dependence of the 
tunneling conductance $G$ 
through a weak link between two Luttinger liquids (LL)
or between a Luttinger liquid and a normal metal (FL)
\begin{equation}
G(T) = G_{\Psi}^{\rm left}(T) G_{\Psi}^{\rm right}(T) \sim 
\left\{ 
\begin{array}{l}
T^{4 x_{\Psi,b} - 2 } = T^{\frac{1}{\krm} + \frac{1}{K_{\sigma}} - 2}
\;, \quad (LL-LL) \\
T^{2 x_{\Psi,b} - 1 } = T^{\frac{1}{2 \krm} + \frac{1}{2 K_{\sigma}} - 1}
\;, \quad (LL-FL) 
\end{array}
\right.
\end{equation}
In magnetic fields, and for OBC, 
the exponents controlling these temperature dependences depend \em
in  first order \rm on $\ln(h_c/h)$, unlike periodic systems where
the dominant dependence of the Green's function exponent is of second
order in $\ln(h_c/h)$. This also holds for the related situations where
particles scatter off a weak impurity in a periodic
ring \cite{kafi} because here, the temperature dependence of the conductance
is controlled by the $2k_F$-part of the density correlations, 
\begin{equation}
G(T) = G_0 \left[ 1 - {\rm const.} \; 
T^{K_{\rho} + K_{\sigma} -2 } \right] \;.
\end{equation}
Here, a finite $h$ will increase the exponent while in the tunneling case,
the exponent will be decreased. 
If $K_{\rho} < 1$, in both cases the effect of a finite magnetic field
will be to offset the interaction effects on the exponents contained in
$K_{\rho}$, and to produce conductance variations 
more similar to free electrons. Precisely this effect has apparently
been observed in semiconductor quantum wires in magnetic fields 
\cite{rother}. Notice that in such structures, $E_F$ is small, typically
some $meV$. Consequently, $h_c$ is also small and the magnetic field
effect on $K_{\sigma}$ could well be big. 

The density-density correlation function 
\begin{equation}
G_{n}(x_1,x_2,t)= \langle n({x_1},t)n({x_2},0) \rangle \; ,
\end{equation} 
involves the following excitations
\begin{eqnarray}
\Delta N_c & =& \Delta N_s=0, {~~~}\Delta I_c=1,{~~~} 
\Delta I_s=0 {~~~}{\rm or}
{~~~} \Delta I_c=0,{~~~} \Delta I_s=1 \;, \\
\Delta N_{\rho} & = & 0, \quad \Delta N_{\sigma} = 0, \quad \Delta I_{\rho}
= 1, \quad \Delta I_{\sigma} = 0 \;.
\end{eqnarray}
Notice that for open boundary conditions, there are no current excitations
while in the bulk, the density operator can involve current excitations and
therefore oscillate with wavevectors that are even multiples of $k_F$. 
Consequently, there
is a single time behavior while the spatial dependence contains the
usual $q \approx 0, 2k_F, 4k_F, \ldots$ components. We have
\begin{equation}
x_{n,b} =1
\end{equation}
and
\begin{eqnarray}
G_n(0,0,t) & \sim & t^{-2}  \;,\\
G_n(x,0,0) & \sim & x^{-2} + \sin (2k_F x) x^{-\frac{3+ K_{\rho}}{2}}
+ \sin (4k_Fx) x^{-1-2 K_{\rho}} \;\;.
\end{eqnarray}
In the bounded system, there is now a \em universal \rm 
boundary critical exponent $x_{n,b} = 1$, while the time dependence
was non-universal in the periodic system. 
The nonuniversal decay in $x$-direction, and the differences in the
exponents of the various momentum components, 
are a consequence of the different
bulk conformal dimensions of the $2k_F$-harmonics of the density operator.

Friedel oscillations appear in the system because of the breaking of
translational invariance by the boundaries. They have been studied in
detail by Bed\"{u}rftig \em et al., \rm \cite{bed}. We determine their
spatial decay as \cite{wang}
\begin{equation}
\langle n(x) \rangle = n - A_1 \frac{\sin(2k_Fx)}{x^{\frac{\krm + 1}{2}}} 
- A_2 \frac{\sin(4k_Fx)}{x^{2\krm}}
\end{equation}
where we have used the Luttinger liquid language \cite{eggra}. Notice
that they are governed \em not \rm by boundary critical exponents but
by the bulk exponents of the $2k_F$- and $4k_F$-parts of the density 
operator. The boundary exponents of the density correlations, are 
different from those of the Friedel oscillations, however. DMRG
calculations of the Friedel oscillations
are in impressive agreement with conformal field theory
predictions both for spinless \cite{schmitteck} and spin-1/2 fermion 
systems. 

The longitudinal and transverse spin-spin correlation functions involve
the following excitations, respectively,
\begin{eqnarray}
G_{\sigma}^z(x_1,x_2,t) & = & \langle S^z(x_1,t)S^z(x_2,0) \rangle \; ,\\
S^z(x,t) & = & \frac12[n_{\uparrow}(x,t)-n_{\downarrow}(x,t)],\nonumber\\
\Delta N_c & = & \Delta N_s=0, {~~~}\Delta I_c=1,{~~~} \Delta I_s=0 {~~~}
{\rm or } {~~~} \Delta I_c=0,{~~~} \Delta I_s=1, \nonumber  \\
\Delta N_{\rho} & = & 0, \quad \Delta N_{\sigma} = 0, \quad \Delta I_{\rho}
= 0, \quad \Delta I_{\sigma} = 1 ;, \nonumber \\
G_{\sigma}^\perp(x_1,x_2,t) & = & \langle S^-(x_1,t)S^+(x_2,0) \rangle \; ,\\
S^+(x,t) & = & \Psi_{\uparrow}^\dagger(x,t) \Psi_{\downarrow}(x,t),\nonumber\\
\Delta N_c & = & 0,{~~~} \Delta N_s=1, {~~~}\Delta I_c=\Delta I_s=0,
\nonumber \\
\Delta N_{\rho} & = & 0, \quad \Delta N_{\sigma} = 2, \quad \Delta I_{\rho}
=0, \quad \Delta I_{\sigma} = 0 \;. \nonumber
\end{eqnarray}
In the absence of a magnetic field, they decay with the same exponents
as the long-wavelength and the $2k_F$-components of the density 
correlations. The enhancement of magnetic over density
correlations which we expect in the $U>0$-Hubbard model, in the bulk
is due both to logarithmic corrections and to prefactors \cite{hjs,log}.
A similar enhancement in the \em correlation functions, \rm is expected 
close to the boundary. Notice, however, that the static Friedel oscillations
in the density will dominate there, and the importance of the magnetic
correlations in the system will be limited either to dynamical measurements,
or to the bulk region. It is also important to notice that, in the 
Luttinger liquid picture, it is the special value $K_{\sigma} = 1$,
embodying spin-rotation invariance, which makes the boundary critical
exponents of the longitudinal and transverse spin correlations identical.
Formally, we have 
\begin{equation}
x_{S,b}^{\|} = 1 \quad {\rm and} \quad x_{S,b}^{\perp} = 1 / K_{\sigma} \;.
\end{equation}

Finally pairing correlations are important. 
The singlet pairing correlations are 
\begin{eqnarray}
G_{SS}^{(0)}(x_1,x_2,t) & = & \langle \Psi_{\uparrow}(x_1,t)
\Psi_{\downarrow}(x_1,t)
\Psi_{\downarrow}^\dagger(x_2,0)\Psi_{\uparrow}^\dagger(x_2,0) \rangle \; ,\\
\Delta N_c & = & 2,{~~~} \Delta N_s=1, {~~~}\Delta I_c=\Delta I_s=0.\nonumber
\\
\Delta N_{\rho} & = & 2, \quad \Delta N_{\sigma} = 0, \quad \Delta I_{\nu}
= 0 \;.
\end{eqnarray}
Their boundary critical exponent is 
\begin{equation}
\label{bcesp}
x_{SS,b} = \frac{2}{\xi_0^2(k)} = \frac{1}{K_{\rho}}
\end{equation}
so that they decay as
\begin{eqnarray}
G_{SS}^{(0)}(0,0,t) & \sim & t^{-2/\krm} \\
G_{SS}^{(0)}(x,0,0) & \sim & x^{-\frac{3}{2 \krm} - \frac{1}{2}}
\end{eqnarray}
They will dominate for attractive interactions ($\krm > 1$) only. Then,
however, the presence of a boundary will enhance them over their bulk
values.

Triplet pairing correlations have three components ($S^z = 1, 0, -1)$)
involving the following operators and excitations 
\begin{eqnarray}
G_{TS}^{(1)}(x_1,x_2,t) & = & \langle \Psi_{\uparrow}(x_1+1,t) 
\Psi_{\uparrow}(x_1,t) 
\Psi_{\uparrow}^\dagger(x_2,0)\Psi_{\uparrow}^\dagger(x_2+1,0) \rangle \; ,\\
\Delta N_c & = & 
2,{~~~} \Delta N_s=0, {~~~}\Delta I_c=\Delta I_s=0,\nonumber \\
\Delta N_{\rho} & = & 2, \quad \Delta N_{\sigma} = 2, \quad
\Delta I_{\nu} = 0 \;, \nonumber \\
G_{TS}^{(0)}(x_1,x_2,t) & = & \langle \Psi_{\uparrow}(x_1,t)
\Psi_{\downarrow}(x_1,t)
\Psi_{\downarrow}^\dagger(x_2,0)\Psi_{\uparrow}^\dagger(x_2,0) \rangle \; ,\\
\Delta N_c & = & 2,{~~~} \Delta N_s=1, {~~~}\Delta I_c=\Delta I_s=0,\nonumber
\\
\Delta N_{\rho} & = & 2, \quad \Delta N_{\sigma} = 0, \quad
\Delta I_{\nu} = 0 \;, \nonumber \\
G_{TS}^{(-1)}(x_1,x_2,t) & = & \langle \Psi_{\downarrow}(x_1+1,t) 
\Psi_{\downarrow}(x_1,t) 
\Psi_{\downarrow}^\dagger(x_2,0)\Psi_{\downarrow}^\dagger(x_2+1,0) 
\rangle \; ,\\
\Delta N_c & = & 
2,{~~~} \Delta N_s=2, {~~~}\Delta I_c=\Delta I_s=0.\nonumber \\
\Delta N_{\rho} & = & 2, \quad \Delta N_{\sigma} = - 2, \quad
\Delta I_{\nu} = 0 \;, \nonumber 
\end{eqnarray}
Using the quantum numbers of the excitations, we find for $S^z = \pm 1$,
\begin{equation}
x_{TS,b}^{(\pm 1)} = \frac{2}{\xi^2(k_0)} +1 = \frac{1}{\krm } + 1
\end{equation}
This is different from the value of $S^z = 0$ which is equal to that for
singlet pairing, and given in (\ref{bcesp}). Superficially, this would
imply a breaking of spin-rotational invariance which is inconsistent with
the model Hamiltonian. The solution of the puzzle is contained in 
Eq.\ (\ref{psib}). In a periodic Luttinger liquid, the triplet pairing
operators with spin projection $S_z = +1$ and $0$ are represented as
\begin{equation}
O_{TS,+1}(x) = \Psi_{+,\uparrow}(x) \Psi_{-,\uparrow}(x) \quad {\rm and}
\quad O_{TS,0}(x) = \frac{1}{\sqrt{2}} \left[
\Psi_{+,\uparrow}(x) \Psi_{-,\downarrow}(x) +
\Psi_{+,\downarrow}(x) \Psi_{-,\uparrow}(x) \right] \;.
\end{equation}
The linear dependence of right- and left-moving fermions, Eq.\ (\ref{psib}),
in the presence of open boundaries clearly affects the operator 
$O_{TS,+1}(x)$, and is accounted for correctly by our rules for the
critical exponents. On the other hand, the individual contributions
$\Psi_{+,\uparrow}(x) \Psi_{-,\downarrow}(x)$ to $O_{TS,0}(x)$,
underlying the critical exponent (\ref{bcesp}) would not
seem to be affected by (\ref{psib}) because different spin projections
are involved. However, when adding both terms in $O_{TS,0}(x)$, it is clear
that they add up to zero when using (\ref{psib}). Therefore, the prefactor
of the term in the correlation function $G_{TS}^{(0)}(x_1,x_2,t)$
carrying the boundary exponent (\ref{bcesp}) must vanish, and only the next
higher term survives. We therefore conclude that the boundary critical
exponent of the triplet pairing correlations is
\begin{equation}
x_{TS,b}^{(0)} = x_{TS,b}^{(\pm 1)} = \frac{2}{\xi^2(k_0)} +1 
= \frac{1}{\krm } + 1 \;. 
\end{equation}
The singlet correlations are not affected by this argument and (\ref{bcesp})
continues to hold. Note that $x_{TS,b}  = x_{SS,b} + 1$ while their 
bulk scaling dimensions (and therefore their critical behavior in periodic
systems) are equal.

Gapless 1D quantum system with open boundaries form a new universality
class which we propose to call ``bounded Luttinger liquids''. As 
we have shown earlier \cite{wang}, they are $c=1$ conformal field theories,
or products thereof, and they posses critical exponents different from
bulk Luttinger liquids. With bulk Luttinger liquids, they share the
fact that the scaling relations between the nonuniversal critical exponents
are parametrized by one renormalized coupling constant $K_{\nu}$ per
degree of freedom $\nu$. These coupling constants are characteristic
of the underlying Hamiltonian and the same irrespective of the boundary
conditions. 
Unlike bulk Luttinger liquids, bounded Luttinger liquids have no current
excitations. This gives \em universal \rm boundary critical exponents
to correlation functions of operators involving current excitations in the
bulk, where they had nonuniversal exponents. Examples are the $2k_F$-
and $4k_F$-components of the density correlations. This should also 
apply to the $3k_F$-Green's function which would be another interesting
example that might be worked out. Also, as we have shown, the linear 
dependence of right- and left-moving fermions introduced by the boundary,
may produce different boundary critical exponents for operators which 
have the same bulk scaling dimensions. 

Finally, while charge-spin separation may be present in a bounded Luttinger
liquid, it will be exceedingly difficult to observe it: all excitations
are localized by the boundaries, and there are no propagating solutions.

When expressed in terms of \kr , \ks , our single-particle Green's
function takes
the same form as those found elsewhere \cite{fabgog,johan,matt}. 
The suggestion
that a new universality class is realized by correlated fermions with
open boundaries, was also made by Mattsson \em et al. \cite{matt}. \rm

\section{Application: the photoemission puzzle in the Bechgaard salts}
\label{appl}
Here, we discuss one way in which these results can be put to use.
The family of organic conductors $(TMTSF)_2 X$ (where TMTSF stands for
the molecule tetramethyltetraselenafulvalene, and X typically is $PF_6$,
$ClO_4$, $ReO_4$, etc., the ``Bechgaard salts'') is one of the candidates
for the realization of a Luttinger liquid in quasi-1D electron systems. 
There is general experimental evidence for strong anisotropy of the electronic
properties and strong electron-electron interactions. More specifically,
several experiments have produced evidence in favor of Luttinger liquid
behavior:
\begin{enumerate}
\item The NMR spin-lattice relaxation is consistent with a Luttinger form
$T_1^{-1} \sim T + T^{\krm}$. Values $\krm \sim 0.15$ were
suggested \cite{NMR}. 
\item Optical conductivity does not directly probe the Luttinger liquid
but its leading irrelevant operators which can relax momentum to the lattice.
Assuming that these stem from electron-electron Umklapp scattering in
a quarter-filled band, theory predicts $\sigma(\omega) \sim 
\omega^{16 \krm - 5}$ on the high-frequency tail of a mid-infrared peak
\cite{giam}. Experiments observe power-law behavior in this region, leading
to $\krm \sim 0.23$ \cite{schwartz}. 
\item Under a pressure of 9 kbar, the resistivity perpendicular to the 
chains shows power-law behavior in temperature \cite{moser}. Theoretically,
the exponent is related to that of the single-particle Green's function,
giving values $\krm \sim 0.25 \ldots 0.3$.
\end{enumerate}
Although the values of \kr\ suggested by the experiments do not agree
precisely, the properties probed apparently are consistent with a Luttinger
liquid. There is, however, one type of measurement whose results 
systematically disagree with Luttinger liquid predictions: photoemission.

The spectral function of a Luttinger liquid generically exhibits two
dispersing signals, due to charge-spin separation, whose singularities
depend on \kr , \cite{myreview,myspec,ms}. For $\krm \sim 1/3 \ldots 1/4$,
one expects a divergence dispersing with the charge velocity and an 
exponent $-1/3 \ldots -7/32 \approx -1/4$ 
while the singularity dispersing with the
spin velocity has a smaller exponent $-1/6 \ldots +1/16$. This is not
observed \cite{zwick}. The result of the experiment on $(TMTSF)_2 ClO_4$
can be summarized
briefly: no charge-spin separation, no dispersion, not even low-energy
peaks. Instead, there is a broad peak at -1 eV whose tail
reaches down to zero intensity at the chemical potential \cite{zwick}.
The leading edge of the spectra is \em independent of the wavevector \rm $k$
of the photoelectrons, and well fitted by a power law, $\sim 
(- \omega)^{\alpha}$, with exponents $\alpha \geq 1$ over a wide range of
energies. 
Moreover, the appearence of the momentum-resolved
spectral functions is strikingly similar to that of an earlier
experiment with low (or no) angular resolution \cite{phot}. 

We now discuss the possibility that this may be due to impurities on
the sample surface. The argument involves three main steps which we 
develop in the following: (i) ARPES probes intrinsic properties of the 
surfaces of the Bechgaard salts; (ii) the surface states from which the
photoelectrons are ejected, are localized; (iii) if these states are
described as bounded Luttinger liquids, the $K_{\rho}$-values suggested
by the bulk measurements together with Eq. (\ref{lspec}) 
give a good description of the photoemission spectra. 

Very similar lineshapes are observed in the insulating
compound $(TMTTF)_2 PF_6$ \cite{zwick}
(selenium substituted by sulfur) which is believed to be a 1D Mott insulator. 
In photoemission, the 
important difference to \tmx\ is that a gap of about 100 meV is observed,
consistent with the values of the charge gap deduced from other experiments.
Very new experiments on $(TMTSF)_2 ReO_4$ which undergoes a metal-insulator
transition due to anion ordering, observe lineshapes similar to 
$(TMTSF)_2 ClO_4$ in the metallic phase and to $(TMTTF)_2 PF_6$ in the
insulating phase, and furthermore can monitor the gap opening as a function of
temperature.
Photoemission therefore is sensitive to spectral changes at the Fermi surface,
and observes intrinsic properties of the surfaces of the Bechgaard salts.

The absence of dispersion in the experimental data
then implies that the photoelectrons are ejected
out of localized states. Then, even an \em angle-resolved \rm photoemission
experiment will measure the local density of states, $\rho(x,\omega) = 
\int dk \rho(k,\omega)$, i.e. the momentum-resolved spectral function 
integrated over a wide range of momenta.
The surface sensitivity of photoemission 
spectroscopy restricts this conclusion to the sample surface only. In the
bulk, there may well be the propagating charge and spin excitations which
are believed to underly the experiments discussed above. (Notice, though,
that apart the longitudinal DC resisitivity not discussed here, many
experiments do not require them explicitly to be propagating.) The origin
of the localization at the surface is not clear, although impurities, 
perhaps introduced
by cleaving the samples, and/or the enhanced one-dimensionality, provide 
plausible causes. 

If the
bulk is assumed to be a Luttinger liquid, and if we approximate the impurities
at the surface
by open boundaries \cite{kafi}, then the chains at the surface consist 
of finite segments which 
must be described as bounded Luttinger liquids \em with the same \kr\
as the bulk. \rm Provided the conditions underlying Eqs. (\ref{gtime}) and 
(\ref{lspec}) are fulfilled, the local spectral function is obtained as
\begin{equation}
\label{spectm}
\rho(x, \omega) \sim | \omega |^{\frac{1}{2 \krm} - \frac{1}{2}}
\sim \left\{ \begin{array}{ll}
| \omega |^{1/2} & \krm = 1/2 \\
| \omega | & \krm = 1/3        \\
| \omega |^{3/2} & \krm = 1/4
\end{array} \right.
\end{equation}
The result for $\krm = 1/2$ has been obtained earlier by Eggert \em et al.
\rm \cite{johan}, who already speculated on the possible influence of 
surface impurities. In particular for values $\krm \approx 1/3 \ldots
1/4$ suggested by optics and DC-resistivity, 
the low-frequency photoemission weight is well described by (\ref{spectm})
for a
bounded Luttinger liquid. Notice moreover that, with photoelectrons 
emanating from localized states, the independence of the results of
angular resolution finds a natural explanation. 

The conditions on which Eq. (\ref{spectm}) is based, imply that
most of the ARPES intensity must come from states which are
dominated by boundary effects. This energy range is widest when the impurities
are rather closely spaced. In such a case, the discretization of the 
energy levels may no longer be negligible. However, these discreteness
effects are likely washed out by averaging over many different segments,
and one returns to the effective semi-infinite bounded Luttinger liquid
underlying Eq. (\ref{gtime}), as demonstrated by 
Mattsson \em et al. \rm \cite{matt} in the related problem of one small segment
studied with  a poor 
experimental resolution. A more detailed investigation of the spectral 
properties of a 1D correlated electron system with random impurities
would, however,  be useful. 

We therefore
come up with a picture of the Bechgaard salts as
an inhomogeneous system where the bulk would be
a standard Luttinger liquid with propagating charge and spin excitations,
and impurities localizing these excitations at the surface.

\section{Open questions}
t\label{oq}
As a consequence of broken translational invariance, a new set of 
boundary critical exponents appear in correlated 1D electron systems
close to open boundaries. They define a new universality class which
we proposed to call \em bounded Luttinger liquids. \rm In this paper,
we explored these new scaling relations both by applying boundary
conformal field theory to the 1D Hubbard model, and by using Luttinger
liquid theory. Both methods can be implemented in a very similar way
and, of course, lead to identical conclusions. 

Our study leaves open a number of questions, however. 
We only investigated open boundary conditions. There is the possibility 
that the boundary critical exponents depend on type of boundary, e.g.
additional boundary fields (magnetic or chemical potential, etc.), boundary
operators (e.g. spins, relevant for the Kondo problem \cite{kondo}), 
superconductors, etc.,
and that new scaling relations are associated with them. 
Such a possibility is suggested by a study of a Luttinger liquid coupled
to a superconductor \cite{winkel}
which finds a local density of states $\rho(x, \omega)
\sim |\omega|^{(K_{\rho}-1)/2}$, instead of our 
(\ref{lspec}) and (\ref{altll}). One could then
be left with a whole manifold of universality classes depending 
on the specific type of boundary conditions. The dressed charge matrices,
and the Luttinger liquid coupling constants $K_{\nu}$ will, however, be
independent of these details. 

The bulk - boundary crossover has not been studied systematically yet.
While this crossover has been studied for the \em local \rm single-particle
density of states in a bounded Luttinger liquid by Mattsson \em et al.
\cite{matt}, \rm the important question of how dispersion in a spectral 
function appears when the length of a bounded electron system is varied from 
microscopic to macroscopic scales, has been left open. In the same way, one 
would like to 
understand, in terms of dispersion and critical exponents, how the open
boundary condition effectively emerges in a correlated 1D electron system
with an impurity.  The extension to the spectral properties of 
systems with many random impurities would also be important. 

We used the theory of a bounded Luttinger liquid to discuss the idea
that impurities at the surface of organic conductors may lead to localized
electronic states which would dominate the photoemission properties
of such materials \cite{zwick}. Such an analysis, using the $K_{\rho}$-values
suggested by experiments probing the propagating bulk excitations, gives
a local density of states consistent with experiments. Of course, the
picture of an inhomogeneous material, with localized states at the surface
and propagating states in the bulk, is somewhat speculative. It would
predict, however, a definite dependence of the results of an experiment
on its probing depth. Such dependences should be searched for, and only
if they are found, this idea should be taken serious.
Finally, theoretical information on spectral weight distribution beyond 
the predictions of conformal field theory \cite{bcesch}, 
and a clarification of the 
energy scales on which they are accurate, would be very important --
even for the periodic systems.

\acknowledgements
We wish to acknowledge useful discussions with B. Brendel, S. Eggert,
H. Johannesson, V. Meden, M. Rother, K. Sch\"{o}nhammer and C. A. Stafford. 
J.V. is a Heisenberg fellow Deutsche Forschungsgemeinschaft,
and received additional support from DFG under SFB 279-B4 in 
the early stages of this work.


\begin{references}
\bibitem{11} K. Binder, in \em Phase transition and critical phenomena, 
	\rm vol. {8}, ed. by C. Domb and J. Lebowitz (Academic Press, 
	London, 1983).
\bibitem{BPZ} A. A. Belavin, A. M. Polyakov and A. B. Zamolodchikov, 
	Nucl. Phys. B {\bf 241}, 333 (1984); D. Friedan, Z. Qiu and 
	S. H. Shenker, Phys. Rev. Lett. {\bf52}, 1575 (1984).
\bibitem{ci} J. L. Cardy, J. Phys. A {\bf 17}, L385 (1984);
	H. Bl\"{o}te, J. L. Cardy and P. Nightingale, Phys. Rev. 
	Lett. {\bf 56}, 742 (1986); I. Affleck, ibid. {\bf 56}, 746 (1986).
\bibitem{fk} H. Frahm and V. E. Korepin, Phys. Rev. B {\bf 42}, 10553 (1990). 
\bibitem{myreview} J. Voit, Rep. Prog. Phys. {\bf 58}, 977 (1995), for a
	review on Luttinger liquids and the properties of strongly 
	correlated 1D electron systems.
\bibitem{Haldane} F. D. M. Haldane, J. Phys. C {\bf 14}, 2585 (1981).
\bibitem{map} F. D. M. Haldane, Phys. Rev. Lett. {\bf 45}, 1358 (1980) and 
	Phys. Lett. {\bf 81A}, 153 (1981).
\bibitem{hjs} H. J. Schulz, Phys. Rev. Lett. 
	{\bf 64}, 2831 (1990).
\bibitem{bcft} J. L. Cardy, Nucl. Phys. B {\bf 240} [FS12], 514 (1984).
\bibitem{wang} Y. Wang, J. Voit, and F.-C. Pu, Phys. Rev. B {\bf 54},
	8491 (1996). 
\bibitem{fuji} S. Fujimoto and N. Kawakami, Phys. Rev. B
	{\bf 54}, 5784 (1996).
\bibitem{gau} M. Gaudin, Phys. Rev. A {\bf 4}, 386 (1971).
\bibitem{rba} H. Schulz, J. Phys. C {\bf 18},  581 (1985).
\bibitem{boun} F. Woynarovich, Phys. Lett. {\bf 108A}, 401 (1985);
	A. Foerster and M. Karowski, Nucl. Phys. B {\bf 396}, 611 (1993);
	H. Asakawa and M. Suzuki, J. Phys. A {\bf 29}, 225 (1996);
	G. Bed\"{u}rftig and H. Frahm, J. Phys. A {\bf 30}, 4139 (1997).
\bibitem{fabgog} M. Fabrizio and A. O. Gogolin, Phys. Rev. B {\bf 51}, 
	17827 (1995).
\bibitem{johan} S. Eggert, H. Johannesson, A. Mattsson, Phys. Rev. Lett
	{\bf 76}, 1505 (1996). 
\bibitem{matt} A. E. Mattsson, S. Eggert, and H.
	Johannesson, Phys. Rev. B {\bf 56}, 15615 (1997). 
\bibitem{bed} G. Bed\"{u}rftig, B. Brendel, H. Frahm, and R. M. Noack,
	Phys. Rev. B {\bf 58}, 10225 (1998).
\bibitem{schmitteck} P. Schmitteckert and U. Eckern, Phys. Rev. B 
	{\bf 53}, 15397 (1996). 
\bibitem{affegg} I. Affleck and  S. Eggert, Phys. Rev. B {\bf 46}, 
	10866 (1992).
\bibitem{bcesch} K. Sch\"{o}nhammer, V. Meden, W. Metzner, U. Schollw\"{o}ck,
	and O. Gunnarsson, cond-mat/9903121.
\bibitem{phot} B. Dardel, D. Malterre, M. Grioni, P. Weibel, Y. Baer, 
	J. Voit, and D. J\'{e}r\^{o}me, Europhys. Lett. {\bf 24}, 687 (1993).
\bibitem{zwick} F. Zwick, S. Brown, G. Margaritondo, C. Merlic, 
	M. Onellion, J. Voit, 
	and M. Grioni, Phys. Rev. Lett. {\bf 79}, 3982 (1997).
\bibitem{grioni} M. Grioni and J. Voit, to be published in
	``Electron spectroscopies applied to low-di\-men\-sio\-nal 
	materials'', edited by
	H. Stanberg und H. Hughes, Kluwer Academic Publ. (1999).
\bibitem{kafi} C. L. Kane and M. P. A. Fisher, Phys. Rev. Lett. {\bf 68}, 
	1220 (1992).
\bibitem{rother} M. Rother, \em et al., \rm unpublished. 
\bibitem{bockrath} 
	M. Bockrath, \em et al., \rm Nature {\bf 397}, 598 (1999).
\bibitem{ba} E. H. Lieb and F. Y. Wu, Phys. Rev. Lett. {\bf 20}, 1445 (1968).
\bibitem{charge} We will also use the term ``charge excitation" for particle
	addition, although these may occur both in the charge and spin 
	sectors. From the context, it should be clear if ``charge" refers
	to an excitation, or to a sector of the Hilbert space.
\bibitem{notation} Charge and spin labels are used in a different sense
	in Bethe Ansatz and Luttinger liquid theory. We therefore use
	the subscripts $c$ and $s$ in a Bethe Ansatz context, and the
	greek subscripts $\rho$ and $\sigma$ in a Luttinger liquid context.
	When we refer to the same quantities in both methods (e.g. the
	velocities $v_{\rho}, v_{\sigma}$), we prefer the Luttinger liquid
	notation.
\bibitem{fkmag} H. Frahm and V. E. Korepin, Phys. Rev. B {\bf 43}, 5653
        	(1991).
\bibitem{ksh} I. Affleck and M. Oshikawa, cond-mat/9905002. N. M. Bogoliubov,
	A. G. Izergin, and V. E. Korepin, Nucl. Phys. {\bf B 275}, 687 (1986).
\bibitem{kt} K. Kawano and M. Takahashi, J. Phys. Soc. Jpn. {\bf 64}, 4331 (1995).
\bibitem{hallb} K. Penc, K. Hallberg, F. Mila, and H. Shiba, 
	Phys. Rev. B {\bf 55}, 15475 (1997).
\bibitem{eggra} R. Egger and H. Grabert, Phys. Rev. Lett. {\bf 75},
	3505 (1995). 
\bibitem{log} J. Voit, J. Phys. C {\bf 21}, L-1141 (1988). 
\bibitem{NMR} C. Bourbonnais, F. Creuzet, D. J\'{e}rome, K. Bechgaard,
	and A. Moradpour, J. Phys. (Paris) Lett. {\bf 45}, L-755 (1984); 
	P. Wzietek, F. Creuzet, C. Bourbonnais, D. J\'{e}rome, 
	and A. Moradpour, J. Phys. (Paris) I {\bf 3}, 171 (1993).
\bibitem{giam} T. Giamarchi, Phys. Rev. B {\bf 44}, 2905 (1991).
\bibitem{schwartz} A. Schwartz, M. Dressel, G. Gr\"{u}ner, V. Vescoli,
	L. Degiorgi, and T. Giamarchi,  
	Phys. Rev. B{\bf 58}, 1261 (1998).
\bibitem{moser} J. Moser, M. Gabay, P. Auban-Senzier, D. J\'{e}rome, 
	K. Bechgaard, and J.M. Fabre, Eur. Phys. J. B {\bf 1}, 39 (1998). 
\bibitem{myspec} J. Voit, Phys. Rev. B {\bf 47}, 6740 (1993).
\bibitem{ms} V. Meden and K. Sch\"{o}nhammer, Phys. Rev. B {\bf 46}, 15753 
	(1992).
\bibitem{kondo} This possibility can be investigated in \em special \rm models
	for the Kondo effect in Luttinger liquids which produce 
	``Fermi-liquid-like thermodynamics'', e.g. A. Schiller and K. 
	Ingersent, Phys. Rev. B  {\bf 51}, 4676 (1995); Y. Wang and J. Voit,
	Phys. Rev. Lett. {\bf 77}, 4934 (1996).
\bibitem{winkel} C. Winkelholz, R. Fazio, F. W. J. Hekking, and G. Sch\"{o}n,
	Phys. Rev. Lett. {\bf 77}, 3200 (1996).


\end{references}
\end{document}